\documentclass[10pt,twoside]{article}
\usepackage{Latex-document}
\almvol{00}
\almttone{Frontiers of Science Awards}
\almtttwo{for Math}
\firstpage{1}
\usepackage[english]{babel}
\usepackage[utf8]{inputenc}
\usepackage{multicol}
\usepackage{tabu}
\usepackage{amsmath}
\usepackage{amssymb}
\usepackage{mathscinet}

\usepackage[capitalize]{cleveref}
\usepackage{graphicx}

\numberwithin{equation}{section}

\graphicspath{{./Figures/}}
\newcommand{\st}{\mathsf{t}}

\begin{document}

\markboth{\hfill{\rm Masahito Yamazaki} \hfill}{\hfill {\rm Gauge Theory and Integrability: An Overview \hfill}}

\title{Gauge Theory and Integrability: \\ An Overview}

\author{Masahito Yamazaki}

\begin{abstract}
While general quantum field theories (QFTs) have yet to be rigorously defined in mathematics,
they have generated new mathematics and have served as a unifying principle
connecting different branches of the subject.  In 1989, Witten made a profound impact on the mathematical community 
by systematically constructing 
knot invariants via the three-dimensional Chern-Simons theory.
One of the historical roots of knot invariants was integrable models,
whose explanation in terms of QFT remained unsolved for decades.
Recently, this problem was solved by a perturbative analysis of the four-dimensional
Chern-Simons theory, which provides a novel framework for 
understanding and unifying many different aspects of integrable models.
In this article, we summarize the basic aspects of these developments for non-experts in both physics and mathematics.
\end{abstract}

\maketitle

\setcounter{tocdepth}{1}
\tableofcontents

\section{Mathematical Physics of Quantum Field Theories}\label{sec:intro}

Mathematics is an autonomous discipline that flourishes on the basis of its own problems and motivations. 
At the same time, it has achieved further, sometimes even revolutionary, progress by interacting with the outside world and actively incorporating new ideas.

For the past several decades, Quantum Field Theories (QFTs) have been a tremendous source for novel ideas in mathematics.
While the rigorous mathematical formulation of interacting quantum field theories in four spacetime dimensions has remained an open problem for decades,
we already have a rigorous formulation
for the \textit{perturbative} aspects of QFTs\footnote{What remains unsolved in mathematics is the 
non-perturbative formulation of QFTs. One of the millennium problems of mathematics is to prove the confinement
of the four-dimensional Yang-Mills theory, and this is an example of the problem which requires 
such a non-perturbative formulation.}
(see e.g.\ the books by Costello \cite{MR2778558,MR3586504,MR4300181}).
It has now become possible to study quantum effects of a variety of QFTs
(e.g.\ four-dimensional Yang-Mills theories and their supersymmetric counterparts)
within rigorous mathematics.

In the past, a pure mathematician might have invoked the absence of a rigorous formulation of QFT
as an excuse for not studying the subject. Such an excuse, however, no longer makes sense 
in the 21st century.  In the words of Prof.\ Shing-Tung Yau, 
``In the 21st century, a physicist without mathematical training can hardly become a master, and a mathematician who does not understand theoretical physics has abandoned an important research direction.''\footnote{A quote from the ICBS 2024 conference.}

In this lecture, I introduce the results by 
Costello, Witten and the present author \cite{MR3855889,MR3855890}, wherein perturbative expansions of the four-dimensional
Chern-Simons theory reproduce many aspects of integrable models.\footnote{The YouTube channel by the author
\url{https://www.youtube.com/@masahito.yamazaki} contains some videos related to this article. Readers should also consult the 
summary \cite{Witten:2016spx} by E.~Witten.}
While the article is written primarily in the language of physics, it is also intended for mathematically oriented readers.
Since the four-dimensional Chern-Simons theory may be unfamiliar to many readers,
we begin with a review of the better-known three-dimensional Chern-Simons theory.

 \section{Knot Theory and Three-Dimensional Chern-Simons Theory}\label{sec:knot}

One of the classic results in the application of QFTs to mathematics is Witten's
explanation of knot invariants \cite{MR990772}.

\begin{figure}[htbp]
\centering{\includegraphics[scale=0.25]{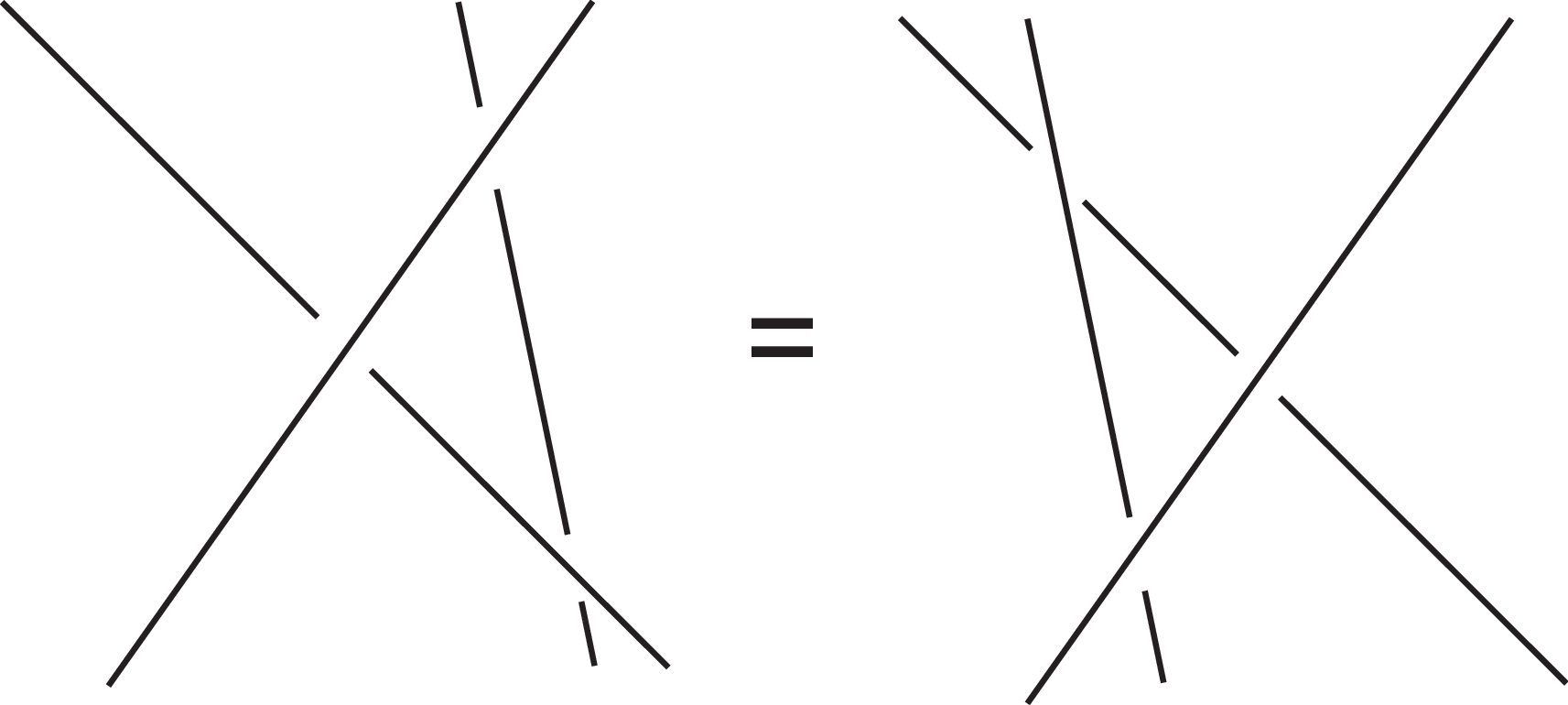}}
\caption{An example of the isotopy of a knot. This is the so-called Reidemeister move III.}
\label{Fig:R_III}
\end{figure}

Consider a knot $K$ inside the three-dimensional sphere $S^3$.
A knot invariant is a quantity $J(K)$ associated with $K$, preserved by continuous deformations (isotopies)
of $K$. In the definition of the Jones invariant by V.~Jones \cite{MR766964,MR908150}
we consider a projection of the knot onto a two-dimensional plane.
While there are some ambiguities in the choice of such a projection
(see \cref{Fig:R_III}), Jones established the well-definedness of the invariant 
by explicitly verifying invariance under changes of projection.

Witten, in contrast, considered the three-dimensional Chern-Simons theory.
Let us choose a semisimple Lie group $G$ and its associated Lie algebra $\mathfrak{g}$.
Let $M$ be an oriented closed 3-manifold with a principal $G$-connection (gauge field) $A$.
The action functional for the three-dimensional Chern-Simons theory is 
\begin{align}
 S[A]= \frac{k}{4\pi} \int_{M} \textrm{Tr}\, \left(A \wedge dA + \frac{2}{3} A\wedge A\wedge A \right) \;,
  \label{eq:3dCS}
\end{align}
where $\textrm{Tr}$ is the Killing form ($\mathfrak{g}$-invariant bilinear pairing)
and $k$ is an integer called the level.
This theory is topological, as the Lagrangian can be defined without the metric on $M$.\footnote{The action functional \eqref{eq:3dCS} may also be written as 
$S[A]= k/(4\pi) \int_{N} \textrm{Tr}\, \left(F \wedge F \right)$,
where we have chosen a four-manifold $N$ whose boundary is $M$ ($M=\partial N$).
This is independent of the choice of $N$ when $k$ is an integer.
Quantum mechanically, the partition function
depends on the signature of $N$ (or the trivialization of $TM\oplus TM$ known as a 2-framing \cite{MR1046621}),
and the theory depends on the choice of the framing in addition to the topology of $M$.
One can obtain a topological invariant of $M$ if one chooses a canonical framing.}
The classical equation of motion gives the flat connection equation $F=dA + A\wedge A=0$, whose solutions
correspond to points of the moduli space of flat connections. The three-dimensional theory gives a procedure for
quantizing such a moduli space.\footnote{In the Hamiltonian quantization
one considers the moduli space of flat connections 
on a two-dimensional time slice $\Sigma$. This is a finite-dimensional symplectic manifold
which can be quantized by methods of geometric quantization.}

Witten considered the Wilson line wherein we integrate the gauge field along the knot $K$:\footnote{By invoking the Borel-Weil-Bott theory,
the Wilson line can be traded for a manifold with boundary $M\backslash N(K)$ \cite{MR990772},
where $N(K)$ is the tubular neighborhood of a knot and we impose appropriate boundary conditions
on the boundary of $M\backslash N(K)$. The equivalence of the two descriptions 
is closely related with a version of the S-duality of the Chern-Simons theory.}
\begin{align}
W_R(K)[A]= \textrm{Tr}_R \, \mathcal{P}\exp \left( \int_K A \right) \;.
\end{align}
Here $\mathcal{P}$ specifies the path-ordering along the knot, and 
the exponential represents the monodromy of the gauge field along the knot.
We have chosen a representation $R$ of $G$, and the trace $\textrm{Tr}_R$ is defined on the weight space
for the representation.

Assume that we have a formal path-integral measure $\mathcal{D} A$ 
in the space of connections modulo gauge transformations.
The expectation value of the Wilson line
\begin{align}
\langle W_R(K)  \rangle = Z^{-1} \int\! \mathcal{D} A \,\, W_R(K)[A]\, e^{i S[A]}  \;,
\quad
Z := \int\!\mathcal{D} A  \,\,  e^{i S[A]} 
\label{eq:WL}
\end{align}
is determined solely by topological data and
does not use a projection of the knot; 
it therefore automatically yields a topological invariant of the knot.

This defines a knot invariant for a given gauge group $G$ and its representation $R$,
and reproduces the Jones invariant for $G=SU(2)$ with $R$ the two-dimensional fundamental representation \cite{MR990772}.
We can moreover obtain invariants of closed 3-manifolds by 
Dehn surgeries along a knot. This procedure can be formulated mathematically e.g.\ 
using the representation theory of the quantum group at a root of unity\cite{MR1091619}.

 \section{Mysteries of Integrable Models}

We now turn to integrable models. 
Historically knots and integrable models have been developing hand in hand,\footnote{I have long been interested
in the connection between the two, ever since I read the review by Wadati et al.\ \cite{MR1015045}.}
and indeed integrable models were among the driving forces behind Jones's discovery of knot invariants.
Here we proceed in the opposite direction: from knots to integrable models.

\begin{figure}[htbp]
\centering{\includegraphics[scale=0.23]{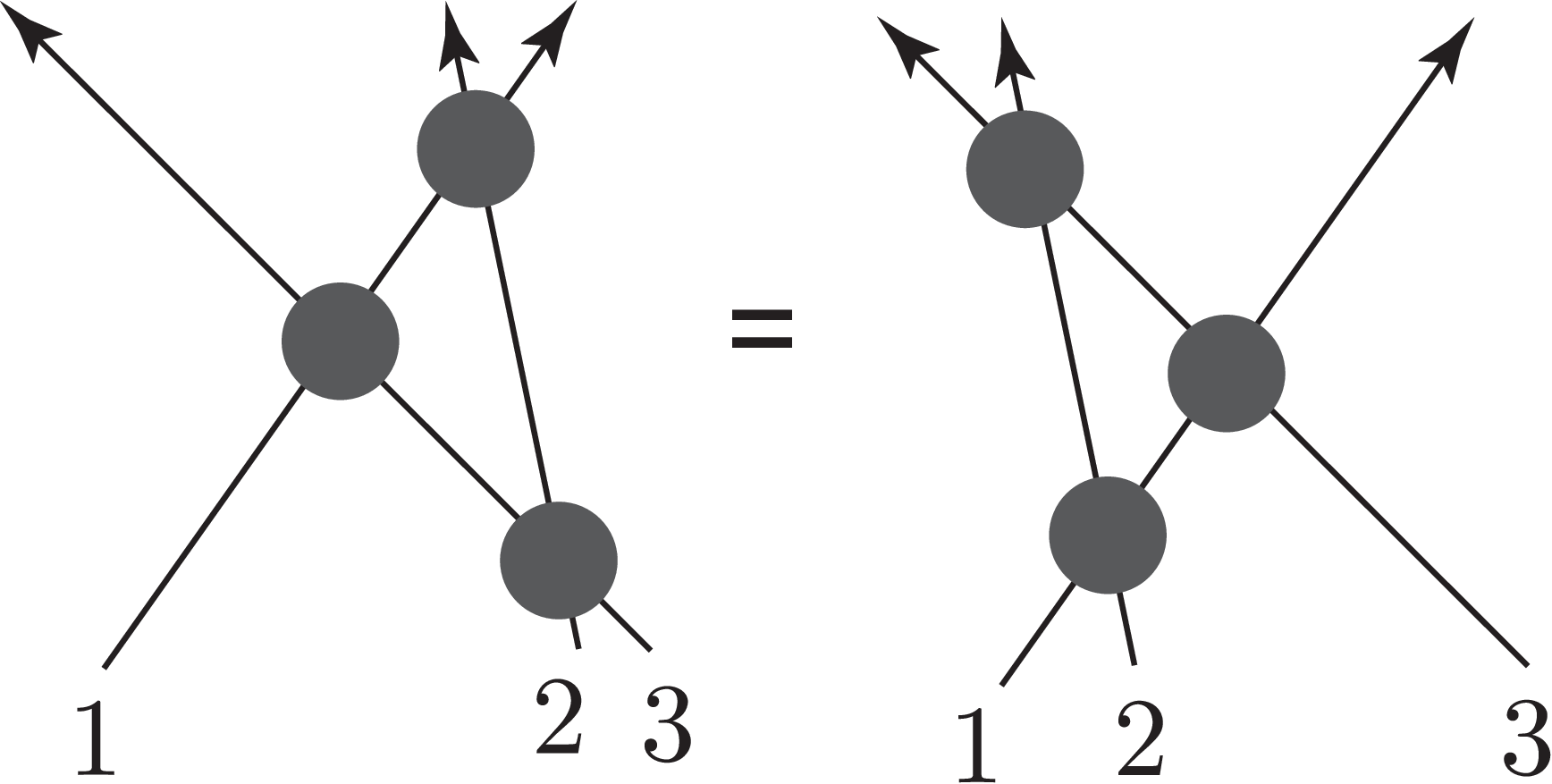}}
\caption{Graphical representation of the YBE.}
\label{Fig:YBE}
\end{figure}

The similarity between knots and integrable models is seen most clearly 
in the graphical representation of the Yang-Baxter equation (YBE) \cite{MR261870,MR290733}
(\cref{Fig:YBE}), whose resemblance to \cref{Fig:R_III} in knot theory is immediately apparent.

\begin{figure}[htbp]
\centering{\includegraphics[scale=0.2]{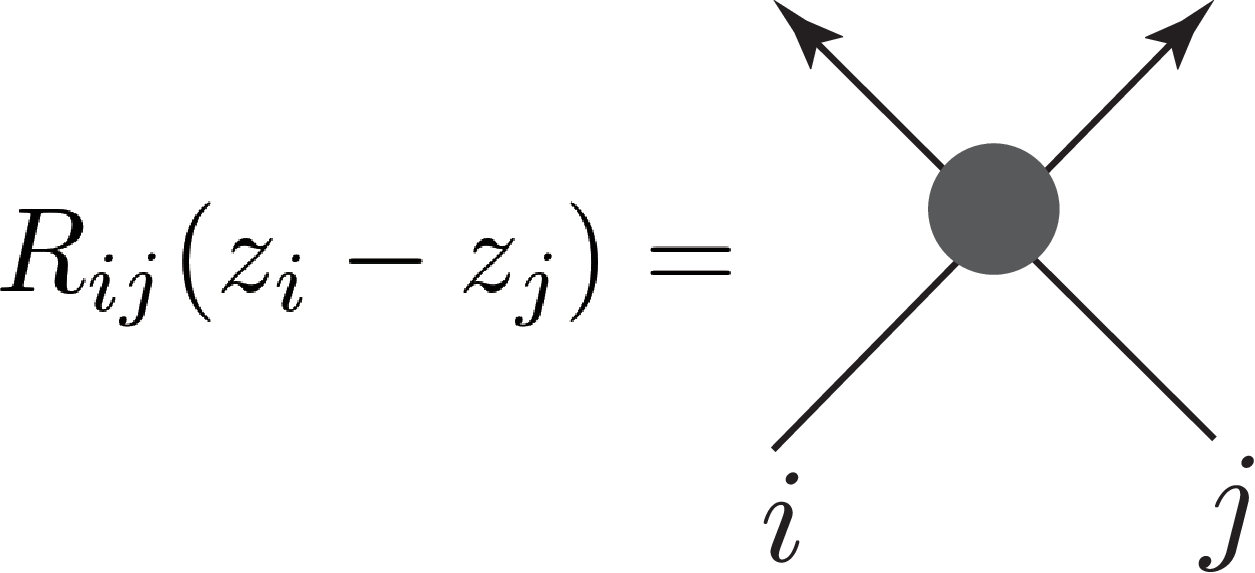}}
\caption{Graphical representation of the $R$-matrix.}
\label{Fig:R-matrix}
\end{figure}

In \cref{Fig:YBE} we have three lines $1, 2, 3$, each 
associated with a vector space $V_{i}$ and spectral parameter $z_{i} \in \mathbb{C}$ ($i=1,2,3$).
At each intersection of two lines lies the $R$-matrix (\cref{Fig:R-matrix})
\begin{align}
R_{ij}(z_i-z_j) \in \textrm{End}(V_i\otimes V_j) \;,
\end{align}
which may be interpreted as describing the scattering of two particles.
The YBE is then an identity between two elements of $\textrm{End}(V_1\otimes V_2\otimes V_3)$:
\begin{align}
\begin{split}
&R_{23}(z_2-z_3) R_{13}(z_1-z_3) R_{12}(z_1-z_2)  \\
& \quad = R_{12}(z_1-z_2)  R_{13}(z_1-z_3) R_{23}(z_2-z_3)  \;,
\end{split}
\label{YBE2}
\end{align}
where $R_{12}\in \textrm{End}(V_1\otimes V_2)$ represents, for example, an element
$R_{12} \otimes 1_{V_3} \in \textrm{End}(V_1\otimes V_2\otimes V_3)$.

The YBE is one of the fundamental characterizations of integrable models,
and its solution defines an integrable lattice model, in which the number of independent conserved charges
equals the number of degrees of freedom.
In this paradigm the classification of integrable models
reduces to the classification of the solutions of the YBE.

The YBE is a highly over-constrained equation, and when one searches blindly, it is not even clear whether any solution exists at all.\footnote{Indeed, most of the physical systems in the real world are non-integrable, and 
there exists a no-go argument for integrable theories in four spacetime dimensions \cite{Coleman:1967ad}. 
One can also add that a non-integrable theory in the UV may flow under the renormalization group to
an integrable fixed-point theory in the IR.} The surprise is that\footnote{While this is often obvious to 
many experts of integrable models, I have an opinion to the contrary. While I may be in the minority,
I am not alone in this respect. For example, Feynman in the late years of his life was interested in integrable models, and tried to derive Baxter's solution to the eight-vertex model \cite{MR290733} in his own way. However, the problem was too difficult and he never succeeded  in solving it \cite{Preskill:2021apy}.}
many different solutions of the YBE have already been constructed in the literature.
Most known solutions fall into the rational, trigonometric and elliptic classification,
where the spectral parameter $z$ takes values in $\mathbb{C},  \mathbb{C}^{\times}=\mathbb{C}\backslash \{0\}$
and the complex one-dimensional torus $E$, respectively.\footnote{The quantum YBE
reduces to the classical YBE when the $R$-matrix has a perturbative expansion in $\hbar$.
One can show, under appropriate conditions, that 
the solutions of the classical YBE are classified into rational, trigonometric and elliptic solutions \cite{MR674005}.
This logic does not apply to some integrable $R$-matrices, e.g.\ those arising from the 
root-of-unity representations of the quantum affine algebra.}
Moreover, there exist Yangians (rational), quantum affine algebras (trigonometric), and elliptic algebras (elliptic),
 which underlie these solutions.
In the following, we focus primarily on the rational case for simplicity.

  \section{Mysteries of the Yangian}

The Yangian, originally introduced by Drinfeld \cite{MR802128,MR914215},
is an infinite-dimensional Hopf algebra whose representation theory 
explains many aspects of integrable models.
While we do not explain its detailed properties (see e.g.\ Chapter 12 of \cite{MR1300632}),  
we note that the Yangian $Y_{\hbar}(\mathfrak{g})$ is associated with a semisimple Lie algebra $\mathfrak{g}$,
and reduces to the universal enveloping algebra $\mathfrak{U}(\mathfrak{g}[\! [z]\! ])$
in the limit $\hbar\to 0$.\footnote{Let us choose a basis $\{\mathsf{t}_a \}$ ($a=1, \dots, \textrm{dim}\, \mathfrak{g}$) of $\mathfrak{g}$,
and their commutators as $[\mathsf{t}_a, \mathsf{t}_b] = \sum_c f_{ab}{}^c \mathsf{t}_c$.
The basis of $\mathfrak{U}(\mathfrak{g}[\! [z]\! ])$ is then given by $\{\mathsf{t}_a^{(n)}= \mathsf{t}_a z^n\}_{n\ge 0}$,
satisfying the commutation relations $[\mathsf{t}_a^{(m)}, \mathsf{t}_b^{(n)}] = \sum_c f_{ab}{}^c \mathsf{t}_c^{(m+n)}$.}

Consider three representations $V_{i}(z_i)$ ($i=1, \dots, 3$) of the Yangian $Y_{\hbar}(\mathfrak{g})$,
where the formal parameters $z_i$ can be identified with the variable $z$ in $\mathfrak{U}(\mathfrak{g}[\! [z]\! ])$.\footnote{Such a representation, labeled by the parameter $z$, is called the evaluation representation.}
To solve the YBE, we can choose the $R$-matrix to be
an intertwiner $R_{ij}(z_i, z_j): V(z_i) \otimes V(z_j)\to  V(z_j) \otimes V(z_i)$
for the tensor product of the two representations---the YBE then follows by 
computing  $V(z_1) \otimes V(z_2)\otimes V(z_3) \to V(z_3) \otimes V(z_2)\otimes  V(z_1)$ in two different methods.

Since the construction of integrable models now follows from the representation theory of the Yangian,
we can automatically obtain solutions of YBE whenever we choose representations.\footnote{We can furthermore construct the universal $R$-matrix,
which specializes to the $R$-matrix upon evaluation in each representation.}
This is of course well-known to the experts on Yangians.
There have been, however, lingering questions for a non-expert like myself.\footnote{Historically, Drinfeld introduced the Yangian by the so-called RTT relation
for the $R$-matrix for the six-vertex model. In other words, to define the Yangian one needs at least one non-trivial solution of the YBE.
If you want to be a minimalist, one should not invoke any $R$-matrix as an input, 
but rather any $R$-matrix should be derived as an output.}
The defining relations of the Yangian
are very complicated (at least to a simple-minded person like the author),\footnote{In mathematics the standard practice when encountering a new definition is to
simply accept it as it is and proceed. This is a practical attitude that I often take.
I feel, however, that I need to internalize the initial definition in my own way to have a proper understanding.
We also note that the defining relations of the algebra become even more involved
as we proceed to trigonometric and elliptic algebras.} and one of the relations for $\hbar\ne 0$ reads as
\begin{align}
\begin{split}
[J(\st_a), J ([\st_b, \st_c]) ]   + [J(\st_b), J ([\st_c, \st_a]) ] + [J(\st_c), J ([\st_a, \st_b]) ]
\\
\qquad \qquad 
=\frac{\hbar^2}{4} \sum_{d,e,f} ([\st_a, \st_d], [[\st_b, \st_e], [\st_c, \st_f]])
\{ \st_d, \st_e, \st_f \} \;,
\label{eq:Y_rel}
\end{split}
\end{align}
where we defined $\{ \st_1, \st_2, \st_3 \} = \sum_{\sigma \in \mathfrak{S}_3} \{ \st_{\sigma(1)}, \st_{\sigma(2)},\st_{\sigma(3)}, \}/ 3!$.
We also defined $\st_a = \st_a^{(0)}$, and $J$ is a map obtained by linearly extending $J(\st_a^{(0)}) = \st_a^{(1)}$.
This equation is nothing but the Jacobi identity when $\hbar=0$, 
and the right-hand side, proportional to $\hbar^2$, represents the quantum correction.
It is natural to ask why the quantum correction takes this particular form.

The relation \eqref{eq:Y_rel} is crucial for the representation theory of the Yangian.
In the classical limit $\hbar=0$, a given representation $R$ of $\mathfrak{g}$
can be naturally extended (by choosing $\st^{(n)}_a=\st_a z^n$) into a 
representation of the Yangian
$Y_{\hbar=0}(\mathfrak{g})=\mathfrak{U}(\mathfrak{g}[\! [z]\! ])$.
The situation is different for $\hbar\ne 0$, where the Yangian relation is 
quantum-corrected and the right-hand side \eqref{eq:Y_rel} 
is the obstruction to the well-definedness of the representation.
For example, the adjoint representation $\mathbf{248}$ of $\mathfrak{e}_8$
is not a representation of the Yangian; it becomes one only after adding
a trivial representation, yielding $\mathbf{248}\oplus \mathbf{1}$ \cite{MR1103907}.
This illustrates the subtlety of the representation theory of the Yangian.

\section{Integrable Models and Four-Dimensional Chern-Simons Theory}

Now that we have explained integrable models and the Yangian,
the next question is how to derive these ingredients from the internal 
logic of QFTs, without appealing to any existing 
results in integrable models per se.

We will be motivated by the similarities between knots and integrable models (\cref{sec:knot}).
Comparing \cref{Fig:R_III} and \cref{Fig:YBE}, we see obvious similarities,
yet two important differences should be kept in mind:
(1) In knot theory we distinguish between over-crossing and under-crossing at the intersection points of the knot projection, and hence 
the picture is intrinsically three-dimensional; in contrast in integrable models we make no such distinctions and the picture is two-dimensional
(2) In integrable models each line has an extra datum of the spectral parameter $z\in \mathbb{C}$.

In standard discussions of integrable models, the statistical lattice is
realized in two dimensions (whose coordinates we denote by $x$ and $y$),
and the spectral parameter is the auxiliary parameter.
Here, by contrast, we incorporate the spectral parameter as part of the geometry, and consider
a four-manifold $\mathbb{R}^2_{x,y} \times \mathbb{C}_z$.\footnote{In integrable models the spectral parameter, which is the argument for the $R$-matrix, sometimes has  an $\hbar$-dependent shift. This is the manifestation of the framing anomaly, which
makes the four-manifold cease to be a direct product in the quantum theory.}
Here $\mathbb{R}^2_{x,y}$ is the two-dimensional space for the integrable lattice,
and $\mathbb{C}_z$ is a complex one-dimensional curve with 
coordinates given by the spectral parameter $z$ and its complex conjugate.
The spectral parameter, which is often treated as an auxiliary parameter, can now be regarded as part of the geometry.\footnote{The idea of geometrization of otherwise non-geometric data in extra dimensions is omnipresent in string theory. This can be regarded as a variant of the Einstein's vision of geometrization of physics.}

We are now ready to write down the action functional for the four-dimensional Chern-Simons theory \cite{Costello:2013zra}:
\begin{align}
  S[A]= \frac{1}{2\pi \hbar} \int_{\mathbb{R}^2_{x,y} \times \mathbb{C}_z} dz\wedge  \textrm{Tr}\left(A \wedge dA + \frac{2}{3} A\wedge A\wedge A \right) \;.
  \label{eq:4dCS}
\end{align}

In the four-dimensional spacetime, the gauge field has four components, 
allowing parallel transport along any of the four directions.
In our case, however, the connection has no $dz$ component since the 
Lagrangian already contains a factor of $dz$, and we obtain the partial connection
$
A = A_x dx + A_y dy + A_{\bar{z}} d\bar{z}
$.\footnote{Related to this point, the action functional is non-Hermitian and 
we need to choose an appropriate integration contour for the convergence of the path integral.
Physically, this is related to the realization of the four-dimensional Chern-Simons theory
as the boundary theory of the partially topologically-twisted 5d $\mathcal{N}=2$ Yang-Mills theory \cite{MR4106881}.
This is analogous to the discussion on the analytically continued three-dimensional Chern-Simons theory \cite{MR2809462}
as the boundary theory for the topologically-twisted four-dimensional $\mathcal{N}=4$ Yang-Mills theory \cite{MR2306566}.}
The classical equations of motion for the Lagrangian \eqref{eq:4dCS}, in the gauge $A_{\bar{z}}=0$,
are: (1) flat in $x, y$-directions, $F_{xy}=0$; (2) holomorphic in the $z$-direction: $\partial_{\bar{z}} A_x=\partial_{\bar{z}} A_y=0$.
The theory is topological along $\Sigma$, and holomorphic along $C$. In particular, the theory
is only partially topological. This is in contrast to the three-dimensional Chern-Simons theory,
which is topological in all three directions.

Since the lattice in integrable models can be regarded as a knot in the three-dimensional Chern-Simons theory,
we consider Wilson lines as in \eqref{eq:WL}:
\begin{align}
W_R = \mathrm{Tr}_R \, \mathcal{P} \, \exp \left( \int_{\gamma} A(z_0)\right) = \mathrm{Tr}_R \, \mathcal{P} \, \exp \left( \int_{\gamma} \sum_a \st_a A^a(z_0)\right) \;.
\label{eq:4dWL}
\end{align}
Since we want to make a lattice along the $\mathbb{R}_{x,y}$-directions, we choose $\gamma$ to lie at $z=z_0$ in $C$, and 
a straight line along either the $x$- or $y$-direction. Inside the trace we consider a representation $R$ of the gauge group $G$.

We are now ready to explain the YBE.
We have the four-dimensional theory with Wilson lines along the topological directions.
The YBE (\cref{Fig:YBE}) states that the physics remains invariant under changes in the relative positions of the Wilson lines,
and this follows immediately in our setup since the theory is topological along the two-dimensional plane.
In other words, this gives a conceptually straightforward explanation of why the YBE admits solutions.
We stress that this argument closely parallels Witten's explanation of knot invariants 
via the three-dimensional Chern-Simons theory.\footnote{One can understand the relation between four-dimensional and three-dimensional Chern-Simons theories as a field-theoretic version of the T-duality in string theory \cite{MR4061201}, wherein the trigonometric quantum affine algebra 
reduces to its subalgebra, the quantum group.
Let us also note in passing that mirror symmetry can be understood as a T-duality \cite{MR1429831},
and the T-duality here can also be regarded as a partial mirror symmetry,
thereby rendering some of the directions of the four-dimensional Chern-Simons theory holomorphic.
The utility of such a ``partial mirror symmetry'' was discussed long ago in the context of 
homological mirror symmetry \cite{MR3100950,MR2423955}.}

\section{Yangian from Quantization}

Let us next discuss in more concrete terms how to obtain Yangians.
The action functional \eqref{eq:4dCS} resembles that of the three-dimensional
Chern-Simons theory \eqref{eq:3dCS} for the $x, y, \bar{z}$-directions, with
$z$ playing the role of an auxiliary parameter.
In other words, the Lie algebra 
$\mathfrak{g}$ is replaced by $\mathfrak{g}[\! [z]\! ]$.
This is consistent with the statement that the classical limit $\hbar=0$
of the Yangian is the universal enveloping algebra $\mathfrak{U}(\mathfrak{g}[\! [z]\! ])$.

A similar observation applies to the Wilson line.
By considering $z$-derivatives of the gauge field, we can generalize \eqref{eq:4dWL} to 
the Wilson line associated with the representation $R$ of $\mathfrak{U}(\mathfrak{g} [\! [z-z_0]\!])$:
\begin{align}
W_R = \textrm{Tr}_{\hat{R}} P \, \exp \left( \int_{\gamma} \sum_a \sum_{n=0}^{\infty} \frac{1}{n!} \underbrace{\st_a (z-z_0)^n}_{\st_a ^{(n)} \in\, \mathfrak{U}(\mathfrak{g} [\! [z-z_0]\!])}
\partial_z^n A^a(z_0)\right) \;.
\label{eq:4dWL_re}
\end{align}

We next consider the quantum effects $\hbar\ne 0$.
While it is true that interacting four-dimensional QFTs have yet to be rigorously formulated, 
we here need only perturbative expansion (with respect to the parameter $\hbar$) of QFTs,
which as mentioned in \cref{sec:intro} can be formulated rigorously in mathematics.\footnote{In classical textbooks on QFTs,
we encounter divergences in the evaluation of Feynman diagrams, which are then ``renormalized'' into parameters of the theory.
Such divergences happen since we integrate up to infinitely high energy scales.
In modern understanding of QFT, however, almost all QFTs are Effective Field Theories (EFTs) and are well-defined only below a certain energy scale (beyond which we expect a different theory, such as string theory). The integral is therefore performed up to a finite cutoff and 
there are no divergences. Most existing attempts for the mathematical formulations of QFTs were formulated before the EFT ideas become popular,
and have not fully incorporated the EFT philosophy.} For perturbative expansion we need to choose a classical solution, for which we choose the trivial solution $A=0$.
We can then expand around this solution, and we can compute the expectation values of observables perturbatively using 
standard QFT techniques.\footnote{The perturbative expansion of the 
three-dimensional Chern-Simons theory was formulated e.g.\ in \cite{MR1225107,MR1341841},
giving rise to finite-type invariants of knots. The Jacobi diagram in this context corresponds to the Feynman diagram of the path integral.}
We can reproduce, for example, the $R$-matrices of integrable models
by perturbative computations for the two Wilson lines crossing each other.
While we will not dwell on the details of the computation here (see \cite{MR3855889,MR3855890} for explicit computations), let us briefly comment on 
how to explain the Yangian relation \eqref{eq:Y_rel} in the perturbative framework.

As already mentioned, the left-hand side of \eqref{eq:Y_rel} represents the Jacobi identity of a Lie algebra.
We may therefore begin by asking why a gauge theory requires a Lie algebra. The answer is provided by consideration of the gauge invariance of the Wilson line 
in the quantum theory.

\begin{figure}[htbp]
\centering{\includegraphics[scale=0.4]{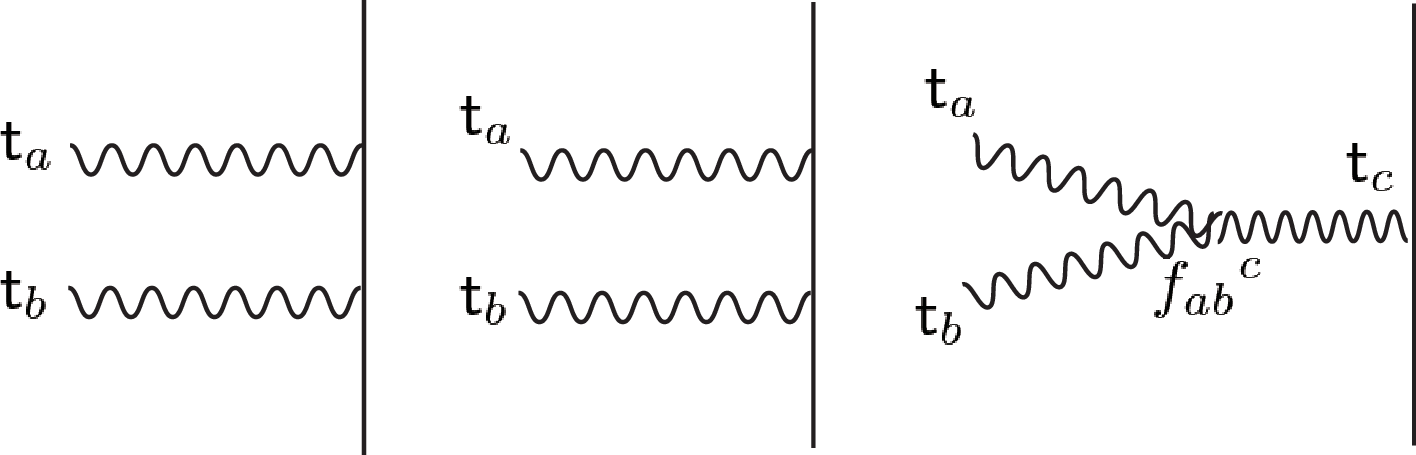}}
\caption{That the gauge field takes values in a Lie algebra follows from the physical requirement that the sum of the three Feynman diagrams be gauge invariant.}
\label{Fig:Jacobi}
\end{figure} 

Let us consider three Feynman diagrams in \cref{Fig:Jacobi}.
Here, a straight line represents the Wilson line associated with a representation $R$ of $\mathfrak{g}$.
A wavy line, on the other hand, represents a gauge field in the adjoint representation of $\mathfrak{g}$, and propagates throughout the four-dimensional spacetime.
These Feynman diagrams have two wavy lines and represent the processes in which two gauge bosons are absorbed or emitted by the Wilson line.
Even without detailed knowledge of Feynman diagrams,
one sees that the three diagrams correspond to the three terms in the commutation relation $\st_a \st_b- \st_b \st_a = \sum_c f_{ab}{}^c \st_c$.
It turns out that each individual Feynman diagram is not gauge invariant, and 
gauge invariance holds only after summing the three diagrams; this is ensured by the commutation relation.
Moreover, by considering the gauge invariance of the Feynman diagrams with three wavy lines, 
one derives the Jacobi identity for the Lie algebra.
This illustrates how we can derive the Lie algebra structure from the QFT.

\begin{figure}[htbp]
\centering{\includegraphics[scale=0.35]{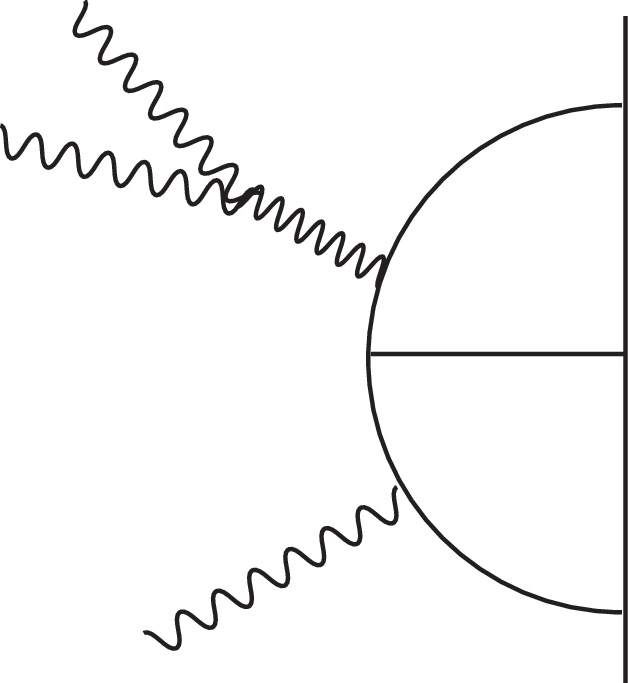}}
\caption{The key Feynman diagram that gives rise to the right-hand side of the Yangian relation \eqref{eq:Y_rel}.
This is a two-loop diagram accounting for the factor of $\hbar^2$ on the right-hand side of \eqref{eq:Y_rel}.
In reality there are many other two-loop Feynman diagrams,
and others must be accounted for by cohomological arguments.}
\label{Fig:2-loop}
\end{figure} 

We now extend the analysis to the quantum theory.
Since the right-hand side of the Yangian relation \eqref{eq:Y_rel} is proportional to $\hbar^2$
we expect corrections at second order in the perturbative expansion.
Indeed, a careful two-loop analysis shows that the 
Feynman diagram of \cref{Fig:2-loop} generates precisely the right-hand side of \eqref{eq:Y_rel},
thus deriving the Yangian relation \eqref{eq:Y_rel} within perturbativation theory.

Note that this argument is valid only up to the second order in the perturbative expansion, and 
does not rule out possible third-order corrections to \eqref{eq:Y_rel}, for example.
To derive the Yangian relation to all orders in perturbation theory
one can appeal to the RTT relation, a variant of the YBE.
This relation can also be justified within the four-dimensional Chern-Simons theory \cite{MR3855890}.

\section{Towards General Quantum Field Theories}

In this article we have highlighted the basic ideas underlying the derivations of 
integrable models from the four-dimensional Chern-Simons theory.
This insight has led to many new results in integrable models,
including the construction of infinitely many new two-dimensional (classically) integrable quantum field theories \cite{Costello:2019tri} (see also \cite{Vicedo:2017cge}).
We can also consider the five-dimensional analog of the Chern-Simons theory in which one direction is topological and the remaining directions are holomorphic,
to derive the affine analog of the Yangians known as affine Yangians \cite{Costello:2016nkh,MR4237705}.\footnote{In recent years Wei Li and the author have identified the quiver Yangians, and more generally, quiver algebras, which are determined not by Dynkin diagram but by a quiver \cite{MR4204250,MR4532676}.
The Chern-Simons interpretation of these algebras is presently unknown, except when the quiver Yangian reduces to an affine Yangian 
associated with the $\mathfrak{gl}_N$ Lie algebra.}

We now summarize the broader lessons from these developments.
The first lesson is that we uncover a wealth of mathematics by studying novel QFTs.
The four-dimensional theory discussed in this paper is
topological along two directions and holomorphic along the remaining two directions,
and hence combines the topological QFT for the former
and the holomorphic part of the two-dimensional conformal field theory (mathematically described by 
vertex operator algebras) for the latter.\footnote{See e.g.\ \cite{Gwilliam:2021zkv} for more general setups.}
Of course, this theory remains a very special example among the vast possibilities of QFTs, and 
one wonders whether entirely new mathematics can be obtained by considering more general theories.
For example, the study of two-dimensional holomorphic field theories 
could lead to higher-dimensional generalization of Kac-Moody algebras \cite{MR3910800,MR4320072}
(see also \cite{Losev:1995cr}).

The second lesson is that seemingly unrelated aspects of integrable models are in fact connected at a deeper level.
For example, a class of equations obtained by the dimensional reduction of the four-dimensional anti-self-dual Yang-Mills
equations has Lax representations and is integrable in some definition \cite{MR1441309}.
This integrability is a priori very different from the integrability as defined by YBE.
There have been attempts, however, to 
connect the two different aspects of integrability by starting from the 
six-dimensional holomorphic Chern-Simons theory \cite{Costello_seminar,Bittleston:2020hfv}.
All these QFTs are realized in superstring theory, in which we can take advantage of the dualities between different theories.
We can therefore say that the contents of this article are all ultimately encapsulated in string theory.
String theory has always unified diverse branches of mathematics and generated unexpected discoveries.
What new mathematics can we extract from string theory, and how it may change mathematics as we know it?
Answering this question remains one of the outstanding questions for twenty-first-century mathematics.

\section*{Acknowledgements}
This article was prepared for the proceedings of the 
International Conference of Basic Science, China, July 2025. Its preliminary Japanese version was 
submitted to the proceedings for the Mathematical Society of Japan, Tokyo, March 2023.
The author would like to thank the organizers and audiences for these events for 
providing stimulating environments.
He would also like to thank Kevin Costello and Edward Witten for 
their excellent collaborations.
This research was supported in part by the
World Premier International Research Center Initiative (WPI), MEXT, Japan, 
by JSPS Grant-in-Aid for Scientific Research (Grant Nos. 23K25865 and 23K17689), 
and by JST, Japan (PRESTO Grant No. JPMJPR225A, Moonshot R\&D
Grant No. JPMJMS2061). 


 \bibliographystyle{ytphys}
 \bibliography{refs}


\address{a) Department of Physics, University of Tokyo, Hongo 7-3-1, Tokyo 113-0033, Japan.\\ 
b) Trans-Scale Quantum Science Institute, University of Tokyo, Hongo 7-3-1,  Tokyo 113-0033, Japan.\\
c) Kavli IPMU, University of Tokyo, Kashiwanoha 5-1-5, Chiba 277-8583, Japan.\\
\email{masahito.yamazaki@ipmu.jp}}

\end{document}